# Direct synthesis of *p*-type bulk BiCuSeO oxyselenides by reactive spark plasma sintering and related thermoelectric properties†

Andrei Novitskii,[a,*] Gabin Guélou,[b,1] Andrei Voronin,[a] Takao Mori[b,c] and Vladimir Khovaylo[a,d]

***Abstract:*** Herein, we demonstrate that BiCuSeO compound can be formed in bulk directly from the raw materials through reactive spark plasma sintering (RSPS) followed by ball milling and a second short spark plasma sintering step. Compared to BiCuSeO samples obtained by a conventional solid-state reaction, the electrical transport properties of the RSPS bulk were moderately affected by the sintering technique, while the lattice thermal conductivity was almost unaffected, and the figure of merit *zT* attained a value at 773 K comparable to state-of-the-art BiCuSeO. The results indicate a new scalable method for the preparation of oxyselenides.

## Introduction

Due to the ability of thermoelectric materials to directly interconvert temperature gradient and electrical power, thermoelectric devices have been recognised as promising candidates for many applications such as waste heat recovery from different energy sources, power generation in deep space or solid-state cooling [1–4]. Performance of the thermoelectric materials can be characterised by the dimensionless figure of merit $zT = \alpha^2 \sigma T \kappa^{-1}$, where $\alpha$ is the Seebeck coefficient, $\sigma$ is the electrical conductivity, $T$ is the absolute temperature, and $\kappa$ is the total thermal conductivity [5]. To achieve a high *zT*, materials are required to have high $\alpha$ and $\sigma$ and low $\kappa$, which are intertwined properties related to the charge carriers and phonons transport. Considering Wiedemann-Franz law and Pisarenko relation [6], the inverse coupling strongly hinders the achievement of high thermoelectric performance [5,7]. Thus, significant attention was received by materials with intrinsically low thermal conductivity, such as Bi-Sb-Te alloys [8], copper sulfides [9,10], borides [11], BiCu*Ch*O (*Ch* = S, Se, Te) oxychalcogenides [12–14] or SnSe [15]. As an emerging layered oxygen-containing thermoelectric material, BiCuSeO was found to exhibit high efficiency in the wide temperature range (*zT* >1.0 at 923 K for Ba doped BiCuSeO) [16,17]. The extremely low thermal conductivity of BiCuSeO (~1 W m$^{-1}$ K$^{-1}$ at room temperature) is believed to originate from large anharmonicity, low Young's modulus, layered crystal structure and large atomic weight of constituents [12,18]. Moreover, the electrical transport properties may be improved by optimizing the charge carrier concentration, tuning the band structure or defect engineering [19–24].

From an industrial point of view, a large *zT* value observed at a given temperature is not the only important parameter. The development of a straightforward scalable synthesis route appropriate for industrial mass production becomes a critical issue for thermoelectric device fabrication. Typically, synthesising polycrystalline BiCuSeO samples is a two-step solid-state reaction route (SSR) involving preparation of feedstock (typically it is already BiCuSeO single-phase) in the form of powder followed by consolidation into a dense sample suitable for transport measurements. SSR technique is cumbersome, time-consuming and energy-intensive. Moreover, it was widely reported that BiCuSeO oxyselenides might be synthesised by various types of powder metallurgy methods, such as mechanical alloying (MA) [25–27] or self-propagating high-temperature synthesis (SHS) [28–30]. However, all techniques include powder fabrication and consolidation steps, while it is favourable to reduce powder handling steps in order to avoid contaminations, partial oxidation etc. In an effort to drive oxyselenides towards another step closer to industrial application, we report on synthesis of BiCuSeO oxyselenides in a bulk form using two-step reactive spark plasma sintering (RSPS).

## Experimental details

The "BiCuSeO" stoichiometric mixture of precursors, Bi, Cu and Bi$_2$O$_3$ powders (> 99.95%) and Se (99.997%), was mixed and ball-milled using a Pulverisette 7 planetary micro mill (Fritsch, Germany) with zirconium oxide balls (Ø = 5 mm, powder-to-ball ratio of 1:5) and bowls (45 ml) in air at 850 rpm for 5 min. The samples were sintered from the powder using a Dr.Sinter-1080 SPS system (Fuji-SPS, Japan) at different temperatures (see Table 1) for 5 minutes under a uniaxial pressure of 50 MPa with heating and cooling rates of 50 K min$^{-1}$ and 20 K min$^{-1}$ respectively. All densified disk-shaped specimens had a dimension of 10 mm diameter × 10 mm height. Noteworthy, the pellets obtained using this synthesis route are generally poorly densified (less than 90% of the theoretical density, see Table 1); to prepare a well-densified sample, a further densification step is necessary; obtained bulk sample were thus ball milled and densified once again at 973 K for 5 min and 50 MPa (labelled as RS7, see Table 1). The as-sintered bulks were annealed at

*a National University of Science and Technology MISIS, Leninsky prospekt 4, Moscow 119049, Russian Federation*
*E-mail: novitskiy@misis.ru*
*b National Institute for Materials Science (NIMS), International Center for Materials Nanoarchitectonics (WPI-MANA), Namiki 1-1, Tsukuba 305-0044, Japan*
*c University of Tsukuba, Graduate School of Pure and Applied Sciences, Tennoudai 1-1-1, Tsukuba 305-8671, Japan*
*d National Research South Ural State University, Lenin prospekt 76, Chelyabinsk 454080, Russia*
*1 Present address: CRISMAT, CNRS, Normandie University, ENSICAEN, UNICAEN, 14000 Caen, France*
*† Electronic Supplementary Information (ESI) available*

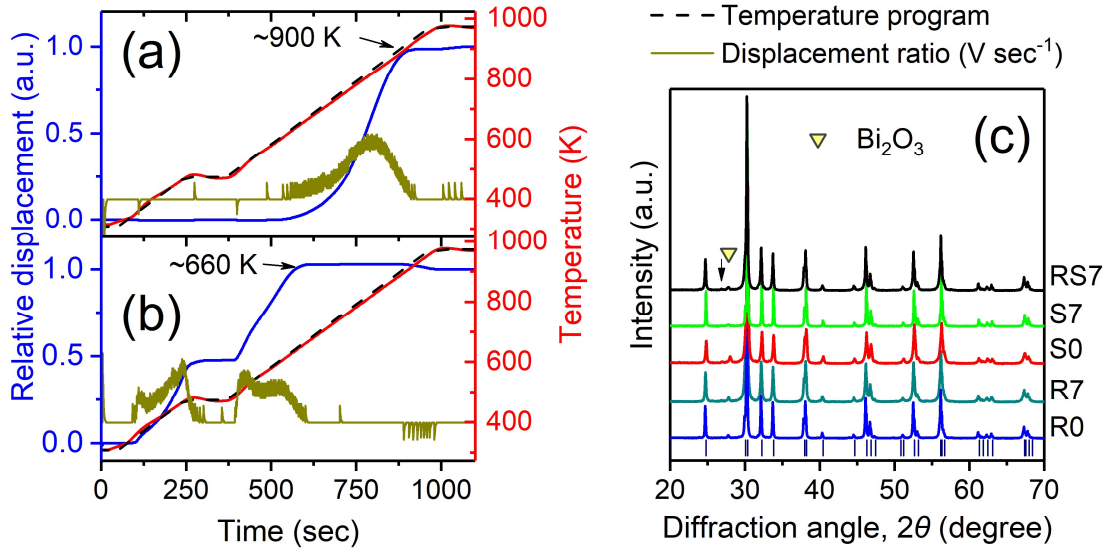

**Figure 1.** The SPS profiles for (a) single-phase BiCuSeO powder during SPS and (b) a stoichiometric mixture of precursors Bi2O3, Bi, Cu and Se during RSPS: relative pistol displacement (shrinkage; blue line), temperature (red line) and displacement ratio (dark yellow line). (c) XRD patterns for BiCuSeO samples after sintering.

**Table 1.** Code, method, sintering temperature, $T_s$, relative volume density, $\rho$, elemental ratios obtained by EDS and lattice constants of the BiCuSeO samples prepared by various methods

| Code | Method | $T_s$ (K) | $\rho \pm 1\%$ (%) | Elemental ratios | Lattice constants (Å) | |
|------|--------|-----------|--------------------|------------------|------------------------|---|
| | | | | Bi:Cu:Se | $a, b$ | $c$ |
| R0 | RSPS | 903 | 89 | 1.02:1.03:0.99 | 3.93004(5) | 8.9301(2) |
| R7 | RSPS | 973 | 87 | 1.03:0.96:1.01 | 3.93166(9) | 8.9339(9) |
| S0 | SSR + SPS | 903 | 94 | 1.02:0.99:0.98 | 3.93176(9) | 8.9308(9) |
| S7 | SSR + SPS | 973 | 94 | 1.04:1.00:0.96 | 3.93007(9) | 8.9314(3) |
| RS7 | RSPS + SPS | 973 | 94 | 1.04:0.99:0.97 | 3.92967(9) | 8.9270(4) |

873 K for 12 hours under Ar atmosphere. For comparison, two more samples were synthesised using a two-step solid-state reaction similar to a previous report [31]. More details on the experimental procedure can be found in the electronic supporting information file (ESI; Fig. S1).

X-ray diffraction data were collected by a SmartLab diffractometer (Rigaku, Japan) using CuKα radiation ($\lambda = 1.5419$ Å) at room temperature. The Rietveld refinements were performed using the FullProf Suite (Fig. S2) [32]. The microstructure of as-synthesised specimens was examined by scanning electron microscopy (SEM; Vega 3 SB, Tescan, Czech Republic). The chemical compositions were analysed by energy-dispersive X-ray spectroscopy (EDS; x-act, Oxford Instruments, UK). Transport properties characterisation was carried out in a direction perpendicular to the pressing direction. The electrical conductivity and Seebeck coefficient were measured simultaneously on bars $1 \times 3 \times 10$ mm$^3$ by the four-probe and differential methods using laboratory-made system (Cryotel Ltd., Russia) under He atmosphere. The thermal diffusivity was measured using a laser flash method (LFA 457, Netzsch, Germany) under continuous Ar flow; the pellets were covered by a thin layer of graphite. The total thermal conductivity was calculated by using $\kappa = \chi \cdot C_p \cdot \rho$, where $\chi$ is the thermal diffusivity, the density, $\rho$, was measured by the Archimedes method and shown in Table 1, and the specific heat capacity, $C_p$, was calculated by the Debye model (see Fig. S3) [33]. Thermogravimetric (TG) analysis was carried out at a heating rate of 10 K min$^{-1}$ using a thermal analyser SDT Q600 (TA Instruments, USA) from room temperature to 1023 K in Ar or air atmosphere. The uncertainty of the Seebeck coefficient and the electrical conductivity measurements was within 8%, and that of thermal conductivity was estimated to be within 5%, considering the uncertainties for $\chi$ and $\rho$. The combined uncertainty for all measurements involved in $zT$ determination was expected to be ~20%.

## Results and discussion

In order to determine a suitable sintering temperature, the SPS profiles of BiCuSeO powder fabricated by SSR and sintered at 903 K, 973 K and 1023 K were analysed. According to the obtained data, the densification process is finished at $T$ ~900K (see Fig. 1a), and any higher temperature should be enough for the fabrication of dense

BiCuSeO sample. However, sintering at 1023 K induced a local liquid phase formation and thereby led to high internal stresses that broke the samples. Thus, for RSPS, the sintering temperatures of 903 K and 973 K were used. SPS profiles for the powders of BiCuSeO phase (Fig. 1a) and the stoichiometric mixture of precursors (Fig. 1b) were completely different. For SSR BiCuSeO powder the SPS profile looked similar to typical sintering process profile for single-phase powders with gradual densification at moderate temperature. On the other hand, for RSPS process the densification rapidly started at $T \sim 350$ K and already reached $\sim 50\%$ at 473 K, where the additional holding step was added; with further temperature increasing the densification process was fast and finished at $\sim 660$ K. Such densification behaviour for raw mixture can be related to the formation of phases occurring along with densification. It can be speculated that the BiCuSeO phase formation path during RSPS was the same as one observed during the SHS process due to their similar mechanisms [29]. However, further comprehensive investigations on this system are necessary in order to understand the phase formation mechanism during RSPS in more details. Considering that the relative density of the RSPS specimens was lower than 90% of theoretical one (see Table 1), the numerous attempts to synthesise dense bulk BiCuSeO samples by one-step RSPS technique using different heating rates, applied pressure and sintering temperature were performed (see Table S1). These attempts were unsuccessful, and the origin of this anomaly is not clear. It can be suggested that induced porosity of the samples may be related to inhomogeneous phase formation or the evaporation of volatile elements during RSPS process.

While single-phase BiCuSeO can be directly prepared from mechanical alloying [25–27], the very short (only 5

minutes) ball milling step was used for the preparation of our powders before RSPS as an efficient way to mix the starting materials homogeneously. It is important to mention that no phases were formed during the ball milling before RSPS and only the precursors' phases were observed in the ball-milled mixture as is shown in Fig. S4. After RSPS, all samples contain mainly one phase, which can be indexed as BiCuSeO (PDF#45-0296) adopting layered tetragonal ZrSiCuAs-type structure with *P4/nmm* space group (Fig. 1c). However, negligible traces of $Bi_2O_3$ and $Cu_{1.8}Se$ impurity phases were detected in all the specimens except for the S7 sample (see labels in Table 1) which is consistent with EDS measurements. The observation of such impurity phases in the composition may be a sign of the presence of copper or bismuth vacancies, which undoubtedly affect the electronic transport of BiCuSeO, as previously reported [34–36]. The unit cell parameters for all the samples are similar and in good agreement with those reported in the literature [29,37]. The EDS measurements indicated that the specimens of the series are homogeneous and the relative elemental proportion of Bi:Cu:Se found from the EDS were close to nominal one within the uncertainty of EDS analysis (see Table 1).

Thermogravimetry measurements of the BiCuSeO samples were performed on bulk pieces under Ar and air atmosphere, respectively (Fig. S5). TG curves for all the samples displayed a similar trend with temperature up to 800 K with a weight loss of less than 0.2% under both air and Ar. However, at a higher temperature, rapid weight loss occurred and the most pronounced increase was observed for the RSPS samples. TG measurements performed under air showed that at $T > 850$ K the SPS and RSPS bulk samples began to oxidise and decompose as observed by C. Barreteau *et al.* [38]. The weight loss can be attributed to Se volatilization at temperatures above 573 K and decomposition of BiCuSeO due to the breakdown of the Cu–Se bonds caused by oxygen from atmosphere entering the crystal lattice [39]. It is worth noting all the samples exhibit better thermal stability over the measured temperature range under Ar flow.

All samples exhibited the same lath-like microstructure with randomly arranged platelet grains, which is typical for oxychalcogenide compounds (Fig. 2). SEM images also displayed the formation of pores with submicron scale sprouting on the surface of the larger particles for R7 sample (indicated by white arrows and dashed ellipse in Fig. 2d), which can be caused by the evaporation of volatile elements as discussed above. However, such pores were observed only for R7 sample and were not the case for R0 sample likely due to its lower sintering temperature.

The electrical transport properties of S7 and RS7 samples were measured for several runs, indicating that RS7 was slightly far from equilibrium state just after the synthesis, while S7 sample was in a stable state and its $\sigma$ and $S$ were not affected by the measurements (Fig. S6). It seems that RS7 specimen, reached the equilibrium state just after a few runs in working conditions as is shown in Fig. S6. The electrical conductivity of all the samples ranged from 20 to 50 $\Omega^{-1}\,cm^{-1}$ and decreased with temperature up to an upturn temperature, $T_t$ (depending on the sample), indicating degenerate behaviour below $T_t$ (Fig. 3). It should be noted that the electrical conductivity of RSPS specimens was lower than that for SPS samples at room temperature, which was most likely caused by induced porosity of RSPS bulks (see Fig. 3a). However, at high temperature, the values obtained for R0 and R7 specimens were quite close to ones obtained for S0 and S7 bulks. Moreover, the electrical conductivity of the RS7 sample was close to those of S7 specimen at $T < 500$ K and even higher at $T > 500$ K. The presence of porosity may be accounted for by the Maxwell–Eucken expression [40]:

$$\sigma_{eff} = \sigma_{porous} \frac{1 - \varphi}{1 + \beta\varphi}, \qquad (1)$$

where $\sigma_{eff}$ is the corrected value of the electrical conductivity for a bulk medium with theoretical density, $\sigma_{porous}$ is the electrical conductivity measured for porous bulk medium, $\varphi$ is the degree of porosity described by a fraction between 0 and 1, and $\beta$ is the correction factor typically with the value of 2 assuming a nearly spherical pore shape. After porosity correction, the room temperature values of the $\sigma$ exhibited ($33 \pm 3$) $\Omega^{-1}$ cm$^{-1}$ for S0, R0 and R7 samples, and ($55 \pm 4$) $\Omega^{-1}$ cm$^{-1}$ for S7 and RS7 bulks, respectively. Such difference may be attributed to the sensitivity of the BiCuSeO electrical transport properties to various type of defects (vacancies, relevant charge state, anti-sites etc.) as was widely reported by other groups [22,36,41–43]. However, further studies should be carried out for a deep understanding of the effect of synthesis technique on both carrier concentration and carrier mobility.

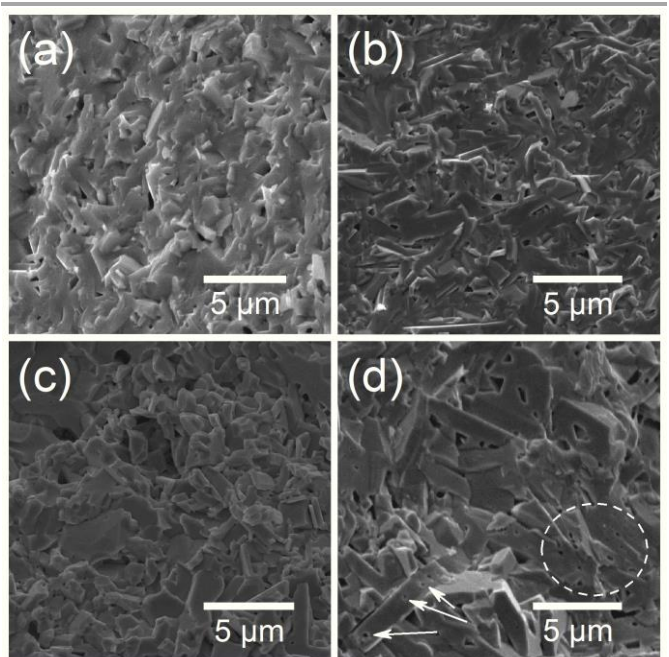

**Figure 2.** SEM images of the fracture surfaces for bulk BiCuSeO fabricated by SPS at (a) 903 K and (b) 973 K, and by RSPS at (c) 903 K and (d) 973 K.

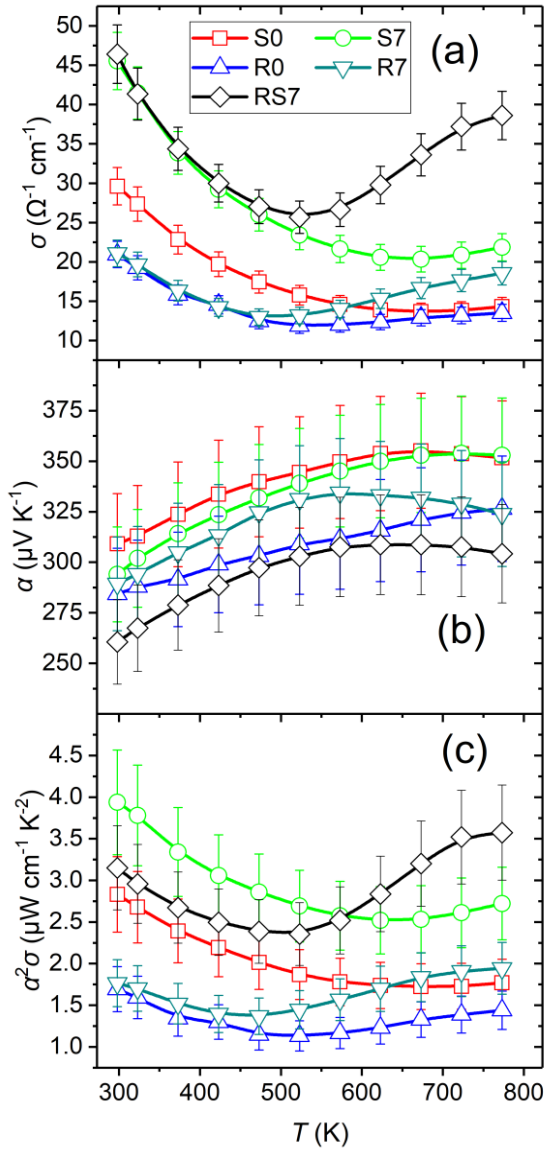

**Figure 3.** Temperature dependence of the (a) electrical conductivity, (b) Seebeck coefficient and (c) power factor for pristine BiCuSeO samples prepared by various methods.

Positive Seebeck values indicated *p*-type conduction with the absolute values of the $\alpha$ increased with increasing temperature (Fig. 3b). The power factor (Fig. 3c) of RSPS samples, due to an induced porosity, was the lowest almost over the whole temperature range. However, mainly due to increase in the electrical conductivity, the power factor of RS7 specimen was the highest at $T > 573$ K with a remarkably high value of ($3.57 \pm 0.57$) $\mu$W cm$^{-1}$ K$^{-2}$ at 773 K for pure BiCuSeO.

The total thermal conductivity data were close to each other for all the samples (Fig. 4a). However, the presence of porosity typically lowers $\kappa$ due to the additional scattering at the pore sites [40,44,45]. Thus, the equation (1) was used to calculate the effective thermal conductivity by replacing the terms $\sigma_{eff}$ and $\sigma_{porous}$ with $\kappa_{eff}$ and $\kappa_{porous}$, respectively [46]. Corrected values of the thermal conductivity exhibited ($1.3 \pm 0.1$) W m$^{-1}$ K$^{-1}$ at room temperature for all the bulks suggesting that $\kappa$ was not significantly affected by the synthesis technique. The lattice thermal conductivity, $\kappa_{lat}$, was estimated by subtracting the electronic contribution, $\kappa_{el}$, from the total thermal conductivity. $\kappa_{el}$ was calculated according to the Wiedemann–Franz law, $\kappa_{el} = \sigma L T$, where $L$ is the Lorenz constant, estimated from the experimental $\alpha$ values using the equation proposed by Kim *et al.*: $L = 1.5 + \exp[-|S|/116]$ (with $L$ in units of $10^{-8}$ W$\Omega$K$^{-2}$ and $S$ in $\mu$VK$^{-1}$) as shown in Fig. S7 [47]. $\kappa_{lat}$ of all the specimens reduced with temperature as $\kappa_{lat} \propto T^{-1}$ (Fig. S8), indicating the dominance of the phonon-phonon scattering, while no obvious bipolar contribution was observed (Fig. 4b). The experimental $\kappa_{lat}$ values of all the studied samples were close to each other over the entire temperature range, reaching ($0.70 \pm 0.06$) W m$^{-1}$ K$^{-1}$ at 773 K. Based on this data, the porosity did not significantly affect the lattice thermal conductivity as it can be expected.

Combining the results of the electrical and the thermal transport properties, the thermoelectric figure of merit $zT$ was calculated and shown in Fig. 4c. The $zT$ values of the BiCuSeO samples followed an increasing trend with temperature (Fig. 4c). A maximum $zT$ was obtained for the RS7 sample, which is comparable and even slightly higher

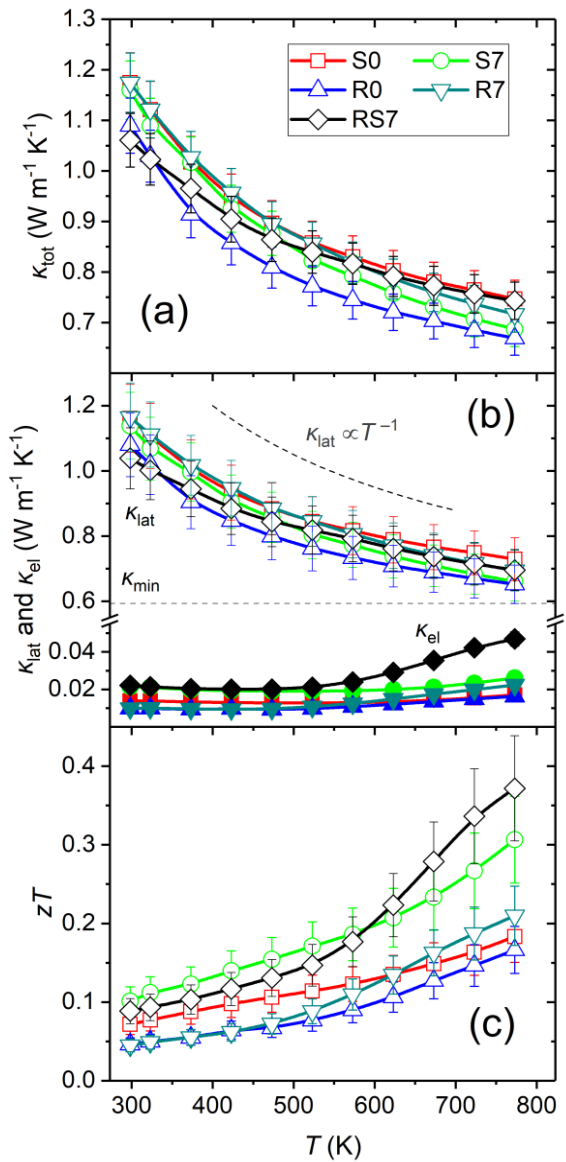

**Figure 4.** Temperature dependence of (a) the total thermal conductivity, (b) the lattice and electronic contribution to the thermal conductivity and (c) the figure of merit $zT$ for pristine BiCuSeO samples prepared by various methods. The minimum lattice thermal conductivity, $\kappa_{min}$, is calculated by Cahill's model [48].

than those for the pristine BiCuSeO fabricated by conventional technique (Figs. 4c and S9).

## Conclusions

In summary, BiCuSeO bulks were successfully synthesised by a one-step RSPS process. However, the density of the obtained bulks was less than 90%, and the origin of such a low density may be attributed to evaporation of volatile elements during sintering. Nevertheless, it was also shown that the use of additional ball milling followed by SPS for RSPS bulks allows one to fabricate dense polycrystalline BiCuSeO bulk. Compared to SSR, the RSPS route can be considered as easily scaled, time and energy-efficient process that allows producing bulks with comparable to state-of-the-art BiCuSeO $zT$ value (Fig. S6). Therefore, further studies aimed at optimisation of RSPS process, as well as the efforts on enhancing thermoelectric performance of the RSPS BiCuSeO based ceramics may be expected in the future.

## Conflicts of interest

The authors declare no competing financial interest.

## Acknowledgements

The study was carried out with financial support from the Russian Science Foundation (project No. 19-79-10282). Vladimir Khovaylo acknowledges Act 211 Government of the Russian Federation (contract № 02.A03.21.0011). Takao Mori acknowledges JSPS KAKENHI JP16H06441 and JST Mirai JPMJMI19A1.

# Direct synthesis of *p*-type bulk BiCuSeO oxyselenides by reactive spark plasma sintering and related thermoelectric properties

Andrei Novitskii,[a,*] Gabin Guélou,[b,1] Andrei Voronin,[a] Takao Mori[b,c] and Vladimir Khovaylo[a,d]

[a] National University of Science and Technology MISIS, Leninsky prospekt 4, Moscow 119049, Russian Federation

[b] National Institute for Materials Science (NIMS), International Center for Materials Nanoarchitectonics (WPI-MANA), Namiki 1-1, Tsukuba 305-0044, Japan

[c] University of Tsukuba, Graduate School of Pure and Applied Sciences, Tennoudai 1-1-1, Tsukuba 305-8671, Japan

[d] National Research South Ural State University, Lenin prospekt 76, Chelyabinsk 454080, Russia

[1] Present address: CRISMAT, CNRS, Normandie University, ENSICAEN, UNICAEN, 14000 Caen, France

[*] E-mail: novitskiy@misis.ru

## Experimental details

The raw powders (ball milled for 5 min, see main text) were put into a cylindrical graphite die and compressed at room temperature for 1 minute under a uniaxial pressure of 50 MPa in an evacuated to ~10 Pa chamber, which then was filled with argon. The temperature of the samples was gradually raised to sintering temperature 903 – 1023 K depending on the sample and additional holding step at 473 K for 2 minutes; after the dwelling for 5 min at a sintering temperature (see Table S1), the pressure was gradually reduced to 30 MPa, and the sample was cooled to room temperature with a cooling rate of 20 K/min. For comparative analysis, BiCuSeO samples were also obtained *via* conventional two-step solid-state reaction (SSR) route using similar to previously reported parameters for powder preparation; prepared powders were densified by spark plasma sintering under the same conditions as in the reactive spark plasma sintering (RSPS) route (903 K or 973 K, 5 min, 50 MPa) [1]. One bulk sample was fabricated by the combination of RSPS and SPS: firstly, the sample was obtained by RSPS at 903 K for 5 minutes under a pressure of 50 MPa; ball milled at 200 rpm for 5 hours and densified again by SPS at 973 K for 5 minutes with an applied uniaxial pressure of 50 MPa (see Table 1 in the main text and Fig. S1). All the densified disk-shaped specimens had a dimension of 10 mm diameter × 10 mm height, respectively.

Figure S1 displays a schematic illustration of the RSPS, SSR + SPS and RSPS + SPS synthesis techniques used in this work. RSPS is the most fast and straightforward route. However, it was not possible to obtain dense samples using only one-step regime (see Table S1). Consequently, so-called two-step RSPS was used for fabrication of BiCuSeO sample with a density of not less than 90%.

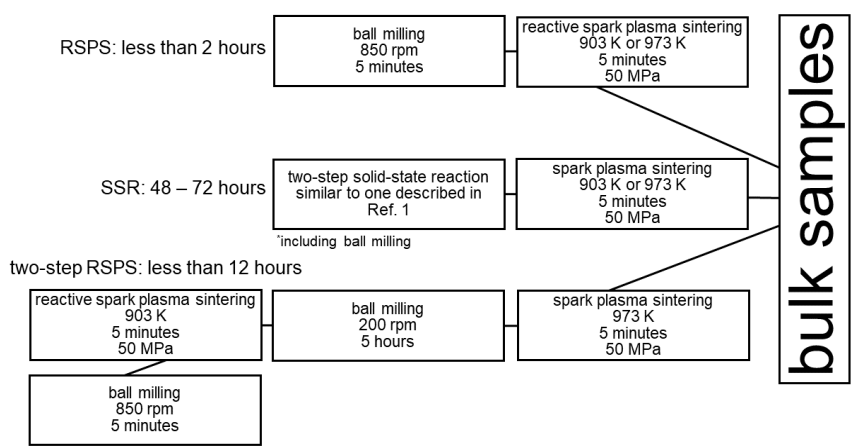

**Fig. S1.** Schematic illustration of the sintering routes for the BiCuSeO samples prepared by various methods.

**Table S1.** Sintering temperature, $T_s$, heating rate, applied pressure and resultant density of the obtained bulk after one-step RSPS sintering.

| $T_s$ (K) | Heating rate (K min⁻¹) | Applied pressure (MPa) | Holding time (min) | $\rho \pm 1\%$ (%) |
|---|---|---|---|---|
| 903 | 50 | 50 | 5 | 89 |
| 973 | 50 | 50 | 5 | 87 |
| 973 | 25 | 50 | 5 | 87 |
| 973 | 50 | 50 | 10 | 89 |
| 973 | 50 | 65 | 5 | 85 |
| 1073 | 50 | 50 | 5 | liquid phase broken sample |

## Rietveld refinement of as-synthesized samples

The final refinement of XRD patterns of as-synthesized specimens was carried out assuming a tetragonal symmetry with a space group of *P4/nmm* and taking the pseudo-Voigt function for the peak profiles. Figure S2 displays the experimental (empty black circles) and simulated (solid red lines) diffraction patterns taken at room temperature, with their differences (solid blue lines) and Bragg positions for phases plotted below the XRD patterns. They agree very well with $R_p \leq 11.9\%$, $R_{wp} \leq 7.7\%$ and $\chi^2 \leq 2.3$.

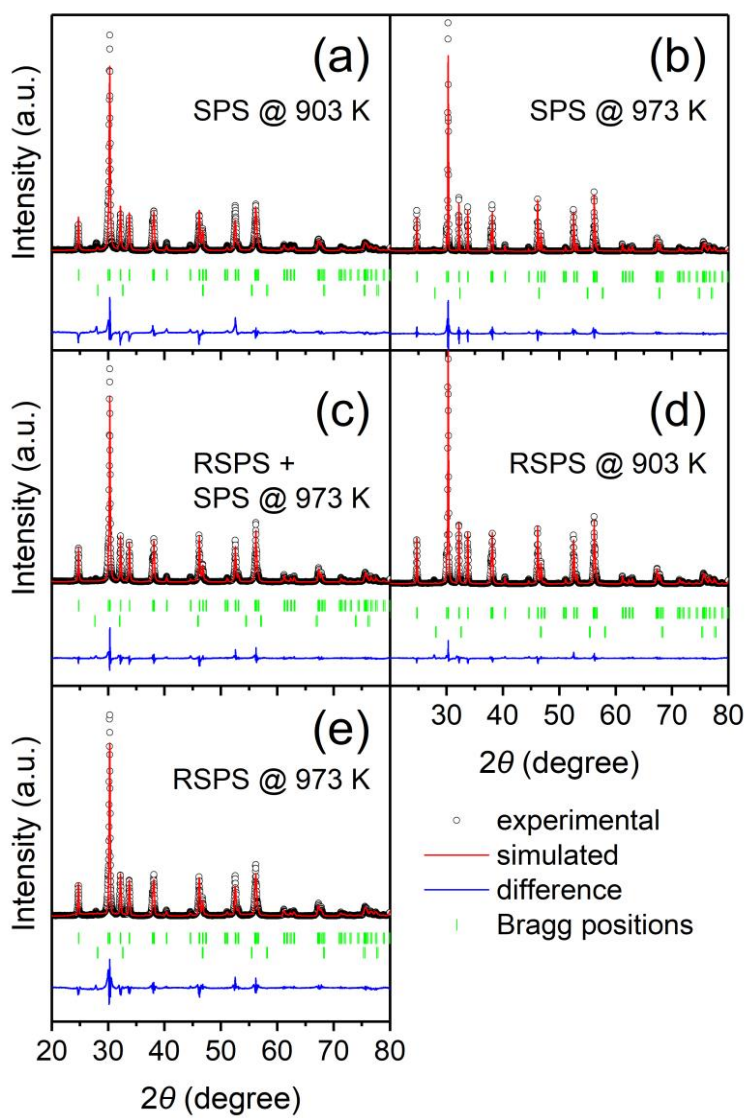

**Fig. S2.** Rietveld refinement of XRD patterns for BiCuSeO samples prepared by various methods: (a) S0, (b) S7, (c) RS7, (d) R0, (e) R7.

## Heat capacity

The specific thermal capacity was calculated by the Debye model (see Fig. S3) [2]:

$$C_p = C_{p(el)} + C_{p(lattice)} = \gamma T + \frac{9Rn}{M} \cdot \left(\frac{T}{\theta_D}\right)^3 \cdot \int_0^{\theta_D/T} \frac{x^4 e^x}{(e^x - 1)^2} dx, \tag{1}$$

where $R$ is the ideal gas constant, $\theta_D$ is the Debye temperature, $n$ is the atom number, $M$ is the molar mass of a primitive cell and $\gamma$ is the Sommerfeld coefficient, $\gamma = 3.075 \cdot 10^{-7}$ J·g$^{-1}$·K$^{-2}$ as reported by several groups [3,4].

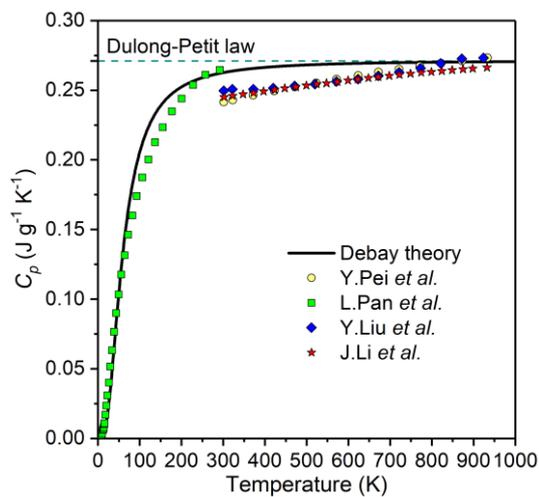

**Fig. S3.** Temperature dependence of the specific heat capacity calculated by the Debye model; some reference data are also shown for comparison [4–7].

**XRD pattern and EDS for raw powder**

To prove, that there were no any reactions or phases formed during ball milling of the raw stoichiometric mixture before RSPS the X-ray diffraction (XRD) analysis and scanning electron microscopy (SEM) were carried out for ball-milled powder (see Fig. S4). According to XRD and energy dispersive X-ray spectroscopy (EDS) analyses, there were no other phases or elements except starting precursors: Cu, Bi, Se and $Bi_2O_3$.

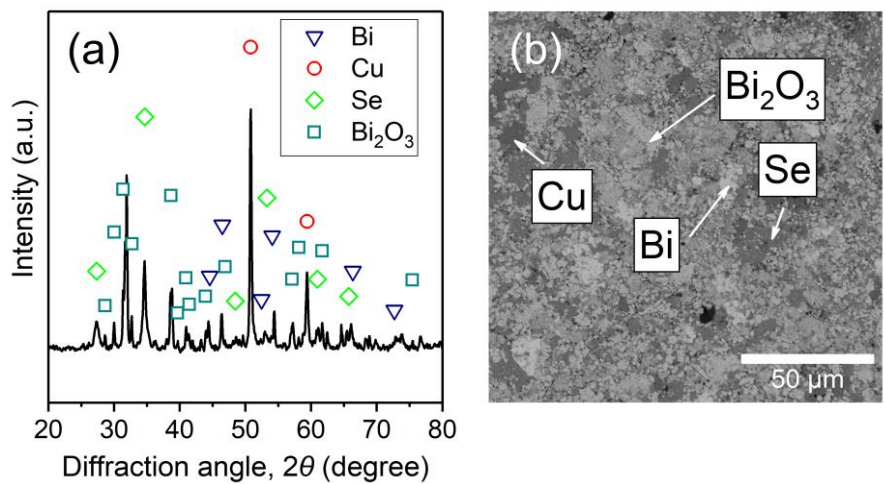

**Fig. S4.** (a) X-ray diffraction pattern and (b) SEM image of the cold-pressed raw stoichiometric mixture after ball milling and before reactive spark plasma sintering.

**Thermogravimetric analysis**

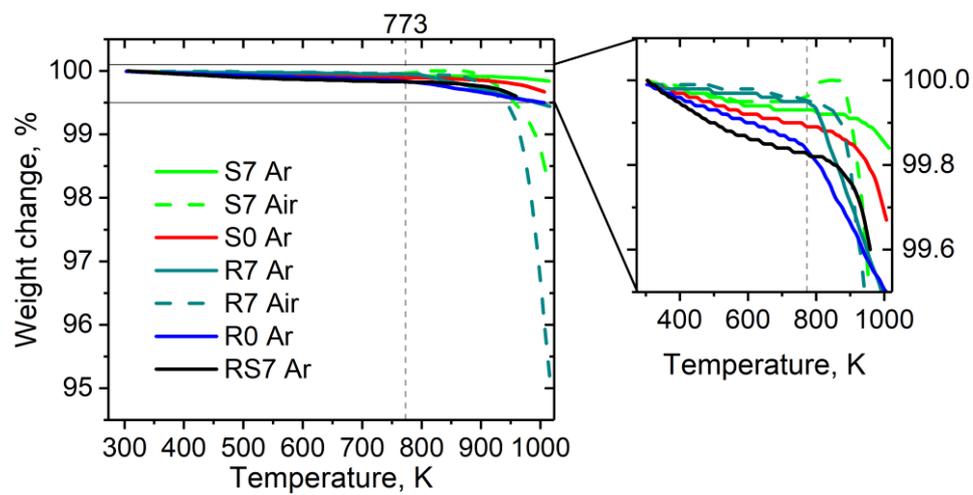

**Fig. S5.** TG curves for BiCuSeO samples prepared by various methods.

**Thermal and electrical transport properties**

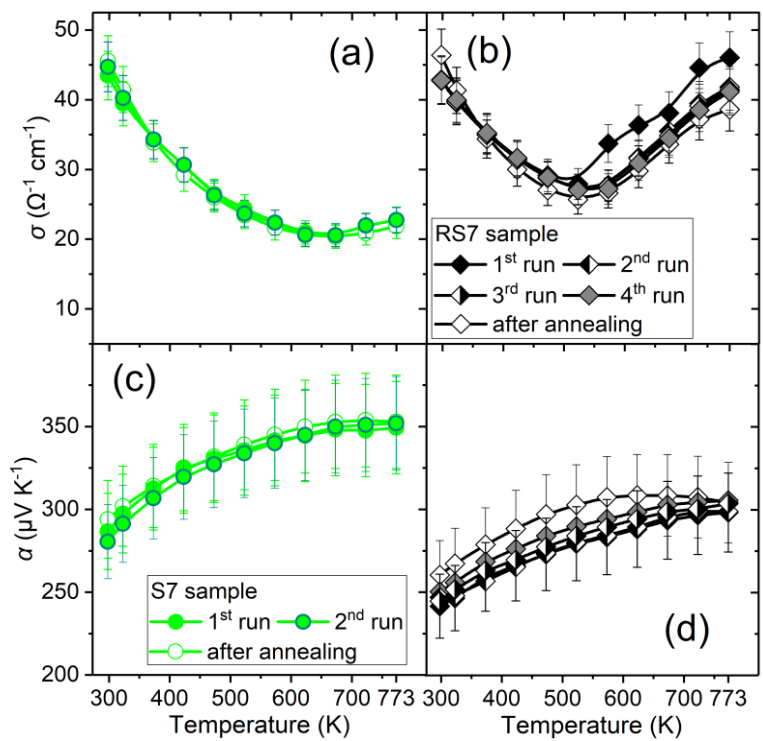

**Fig. S6.** Temperature dependence of (a, b) the electrical conductivity, (c, d) the Seebeck coefficient for S7 and RS7 samples performed for a several runs. Additional annealing for 12 hours at 873 K under argon atmosphere was also performed to compare the stability of the samples.

Lorenz number, $L$, was calculated from the experimental values of the Seebeck coefficient using $L = 1.5 + \exp[-|S|/116]$, where $L$ is in $10^{-8}$ WΩK$^{-2}$ and $S$ in μVK$^{-1}$ [8]. The calculated $L$ of all the samples shown in Figure S7.

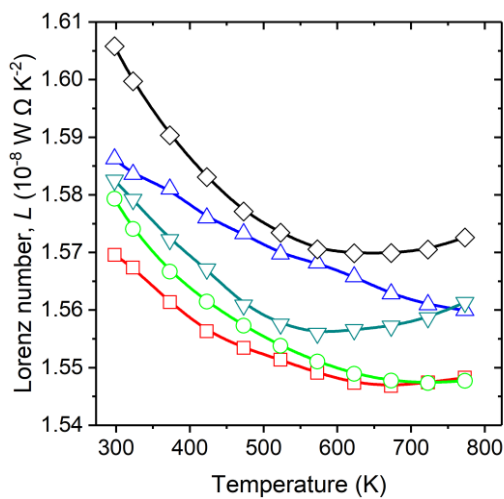

**Fig. S7.** Temperature dependence of the Lorenz number for all the samples prepared by various methods.

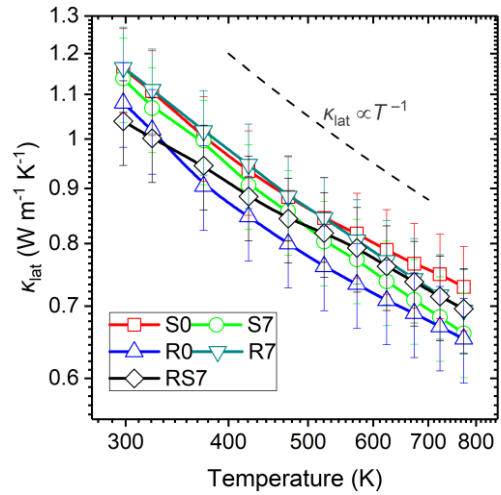

**Fig. S8.** Log-log scale for the temperature dependence of the lattice thermal conductivity of BiCuSeO samples prepared by various methods.

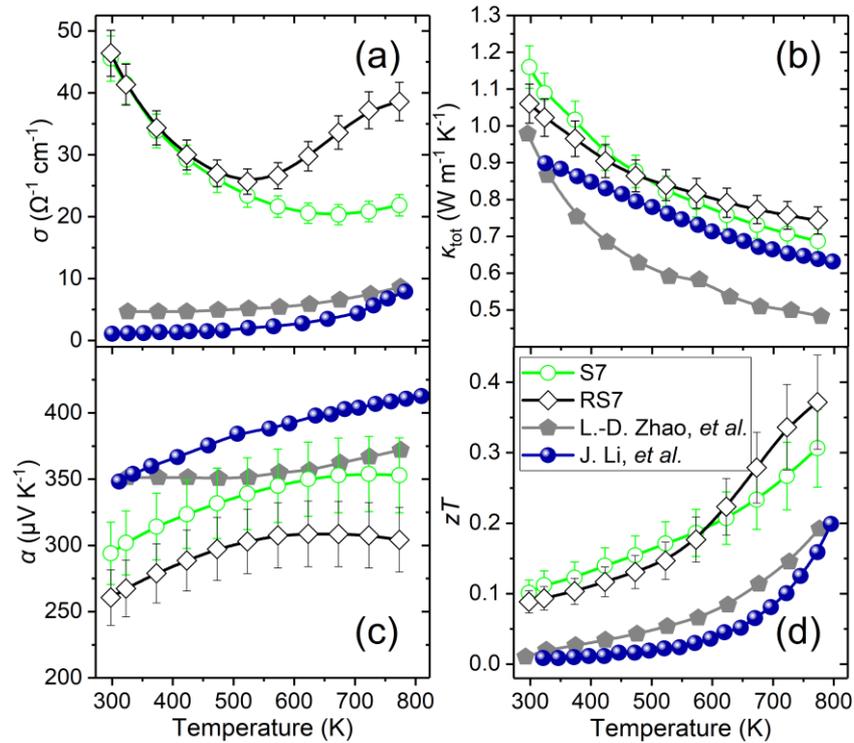

**Fig. S9.** Temperature dependence of (a) the electrical conductivity, (b) the thermal conductivity, (c) the Seebeck coefficient and (d) the figure of merit for S7 and RS7 samples. Data from previous reports also given for comparison [9,10].